\documentclass[aps,twocolumn,showpacs,preprintnumbers,prl,amsmath,amssymb,amsfonts,superscriptaddress,floatfix]{revtex4-1}

\usepackage[paperwidth=210mm,paperheight=297mm,centering,hmargin=2cm,tmargin=1.6cm,bmargin=3.cm]{geometry}
\usepackage{booktabs}
\usepackage{subfigure}
\usepackage{graphicx}
\usepackage{amsfonts}
\usepackage{amssymb}
\usepackage{amsmath}
\usepackage{overpic}
\usepackage{caption}
\usepackage{braket}
\usepackage{xcolor}
\usepackage{cancel}
\usepackage{fancyhdr}
\usepackage[utf8]{inputenc}

\usepackage{dblfloatfix}

\usepackage{bm}
\usepackage{epstopdf}

\newcommand{\be}{\begin{equation}}
\newcommand{\ee}{\end{equation}}
\newcommand{\bea}{\begin{eqnarray}}
\newcommand{\eea}{\end{eqnarray}}

\usepackage{xcolor}
\definecolor{highlight}{rgb}{0, 0.38, 1}
\definecolor{obs}{rgb}{1, 0, 1}
\definecolor{todo}{rgb}{1, 0, 0}

\newcount\bozza \bozza=1
\ifnum\bozza=1
\newdimen\shift \shift=-2truecm

\allowdisplaybreaks

\makeatletter
\makeatletter \renewcommand{\fnum@figure}{{\bf{\figurename~\thefigure}}}
\makeatother

\begin{document}

\title{Prediction of time-reversal-symmetry breaking  fermionic quadrupling   condensate   in twisted bilayer graphene}

\author{I. Maccari}
\affiliation{Department of  Physics, Stockholm University, Stockholm SE-10691, Sweden}
\author{J. Carlstr\"om}
\affiliation{Department of  Physics, Stockholm University, Stockholm SE-10691, Sweden}
\author{E. Babaev}
\affiliation{Department of   Physics, The Royal Institute of Technology, Stockholm SE-10691, Sweden}

\begin{abstract} 
Recent mean-field calculations suggest that the superconducting state of twisted bilayer graphene exhibits either a nematic order or a spontaneous breakdown of the time-reversal symmetry. The two-dimensional character
of the material and the large critical temperature relative to the Fermi energy dictate that the material should have
significant fluctuations. We study the effects of these fluctuations using Monte Carlo simulations. We show that
in a model proposed earlier for twisted bilayer graphene there is a fluctuation-induced phase with quadrupling
fermionic order for all considered parameters. This four-electron condensate, instead of superconductivity, shows
a spontaneous breaking of time-reversal symmetry. Our results suggest that twisted bilayer graphene is an
especially promising platform to study different types of condensates, beyond the pair-condensate paradigm.

\end{abstract}
\maketitle
\noindent
 
 \section{Introduction}
 
 The recently discovered superconducting state which emerges in magic-angle twisted bilayer graphene exhibits a critical temperature that is exceptionally high compared to the Fermi energy
  \cite{caoUnconventionalSuperconductivityMagicangle2018, luSuperconductorsOrbitalMagnets2019,fleurov2019cooperative, yankowitzTuningSuperconductivityTwisted2019, ohEvidenceUnconventionalSuperconductivity2021a}.
This, and the fact that the system is two-dimensional, implies
the presence of strong pairing fluctuations.

While superconductivity is a  more than century-old state of matter, which results from electron pairing, 
the presence of strong fluctuations suggest the tantalizing possibility that magic-angle twisted bilayer graphene can be an especially
promising system to realize different states of matter in the form of condensates of electronic quadruplets.  In principle, the standard Bardeen-Cooper-Schrieffer theory
does not allow fermionic quadrupling condensates. However, if the low-temperature regime of twisted bilayer graphene exhibits a superconducting ground state that breaks multiple symmetries, then, as we show below, it has the ideal ingredients for the formation of fluctuation-induced electron quadrupling states.

Multiple broken symmetries imply a multicomponent order parameter.
Hence, it is described by multiple complex fields of the form $|\Delta_i|e^{i\phi_i}$.
Consider a system that is  a two-dimensional multicomponent superconductor: at finite temperature, and for a finite magnetic-field penetration length, the only 
non-vanishing order parameter 
in the thermodynamic limit has to be constructed out of at least four fermionic fields \cite{babaev2002phase,babaev2004superconductor,Svistunov2015}. 
This is based on the observation that composite superconducting vortices, which have identical phase winding in all components, have finite energy due to supercurrents screening effects. Therefore, a fluctuating two-component system is unstable to the proliferation of composite vortices that disorder the superconducting phase, while preserving the relative density or the phase difference between the components of the order parameter. 
%
The phase difference $\phi_i-\phi_j \propto \arccos \rm{Re} \Delta_i\Delta_j^*$ is an order parameter proportional to the product of two complex fields and hence represents four-fermion correlations.
 Various other realizations of four-fermion order were discussed
 in two-dimensional systems  that exhibit multi-component superconductivity at zero temperature \cite{babaev2002phase,agterberg2008dislocations,berg2009charge,Bojesen2013time,fernandes2021charge,shaffer2021theory,fernandes2021charge,chung2022berezinskii,drouin2022emergent}.

 {The recent microscopic study~\cite{chichinadze2020NematicScTBG} derived an effective mean-field description for twisted bilayer graphene (TBG) near half filling of the valence band ($n=-2$). Upon particle doping, six Van Hove singularities give rise to as many Fermi patches, which are the leading contribution to the density of states. There are two interaction types that are permitted by symmetry, and which contribute to pairing--intra-patch and inter-patch coupling. 
In this scenario, the resulting mean-field theory features two complex order parameters 
$\Delta_1= |\Delta_1| e^{i\phi_1}$ and $\Delta_2= |\Delta_2| e^{i\phi_2}$, and a free-energy potential of the form:}
%
\be
\begin{split}
   V(\Delta_1, \Delta_2)&=  \alpha_1 \left( |\Delta_1|^2 +|\Delta_2|^2\right)  + \\ &+ \beta_1 ( |\Delta_1|^2  + |\Delta_2|^2)^2 + \beta_2\left| \Delta_1^2  + \Delta_2^2 \right|^2,
\label{v1} 
\end{split}
\ee
where $\alpha_1 \propto (T -T_{c_0})$, with $T_{c_0}$ being  the mean field critical temperature, $\beta_1>0$ and $\beta_1 + \beta_2 >0$ for stability. 
The free-energy potential Eq.\eqref{v1}  permits two different ground state manifolds, that are determined by the sign of the coupling $\beta_2$. For $\beta_2>0$, the ground state is a chiral superconductor that breaks time-reversal symmetry, while for  $\beta_2<0$ the superconducting state develops a nematic order. Finally, for $\beta_2=0$, the potential exhibits an $SU(2)$ symmetry.
 { Note that the potential terms of the model are similar to two-component models that appear in many other instances of superconductors that break time-reversal symmetry  \cite{sigrist1991phenomenological}, hence our results apply to other models as well. }

According to the Mermin-Wagner theorem~\cite{Mermin1966},  two-dimensional systems with short-range interactions cannot spontaneously break a continuous symmetry at finite temperatures. 
However, while for the $SU(2)$ symmetry case ($\beta_2=0$), the system does not exhibit any phase transition, the presence of a biquadratic term $\beta_2 > 0$, that explicitly breaks the $SU(2)$ symmetry into a $U(1)\times Z_2$ symmetry, allows for the emergence at low temperatures of an algebraic-ordered superconducting state that additionally breaks a $Z_2$ symmetry. 
When the magnetic field screening is negligible, a two-dimensional system
preserves a $U(1)$ symmetry at any finite temperature~\cite{Mermin1966}, while it exhibits a SC phase transition belonging to the Berezinskii-Kosterlitz-Thouless universality class~\cite{Berezinskii1971, Kosterlitz1972, Kosterlitz1973}.


 {In the limit of strong symmetry breaking, one may consider the London limit and the model can then be mapped onto effective models considered in \cite{Bojesen2013time,Bojesen2014_phase, Haugen2021first,zeng2021phase}.  However, for small $K$, i.e.  in the vicinity of the $SU(2)$ point, fluctuation effects in the density sector may impact the resulting phase diagram.}




In this letter, we focus on the effect of fluctuations in the microscopic model ~\cite{chichinadze2020NematicScTBG} in the scenario $\beta_2>0$.
Starting with the free-energy functional proposed in ~\cite{chichinadze2020NematicScTBG}, we 
employ  large-scale Monte Carlo simulations to obtain the phase diagram of the system beyond the mean-field approximation.

\section{The model}

The Ginzburg-Landau free-energy density of the system reads:

\be
\begin{split}
f&= \sum_{i=1,2 }\left[ \frac{1}{2}| \vec{\nabla}\Delta_i|^2 + \alpha_1 |\Delta_i|^2 \right]  +\\&+ \beta_1 ( |\Delta_1|^2  + |\Delta_2|^2)^2 + \beta_2\left| \Delta_1^2  + \Delta_2^2 \right|^2.
\label{f1}
\end{split}
\ee
When coupled to a gauge field, this model only exhibits a quartic order in the thermodynamic limit~\cite{babaev2002phase}. However, for the case of TBG, the screening is   negligible.
Hence, we consider the  problem of computing the phase diagram in the extreme type-II limit.
The resulting description is characterized by an $SU(2)$ symmetry that is explicitly broken down to $U(1)\times Z_2$ by the $\beta_2$ term. The symmetry-breaking term renders fluctuations of the relative density massive.
These fluctuations can be important in this model 
for the statistical problem of assessing the SC and the $Z_2$ critical temperatures.
Correspondingly, we retain them as part of our description, while taking the total density to be constant, $|\Delta_1|^2 + |\Delta_2|^2=|\Delta_0|^2$.


Rescaling the free energy by the total density $|\Delta_0|^2 = |\alpha_1|/(\beta_1 +\beta_2)$, one can express Eq.\eqref{f1} as a function of a single parameter $K$:

\be
\begin{split}
f= &\frac{1}{2} \left[ |\vec{\nabla}\Delta_1|^2 +|\vec{\nabla}\Delta_2|^2\right]  +\\&+ 2K |\Delta_1|^2|\Delta_2|^2 \left[ \cos(2(\phi_1 - \phi_2)) -1 \right],
\label{f1_rescaled_typeII}
\end{split}
\ee
\\
given by $K= \frac{\beta_2}{\beta_1 +\beta_2} >0$. Next, by collecting the phase-difference gradient terms we obtain the free energy:
\be
\begin{split}
f=  &\frac{1}{2 \rho^2} \left[ |\Delta_1|^2 \vec{\nabla} \phi_1 +  |\Delta_2|^2 \vec{\nabla} \phi_2 \right]^2+ \\& +\frac{|\Delta_1|^2|\Delta_2|^2}{2 \rho^2} \left[ \vec{\nabla} (\phi_1 - \phi_2)  \right]^2  + \\& +\frac{1}{2}\left[ (\vec{\nabla} |\Delta_1|)^2 + (\vec{\nabla} |\Delta_2| )^2\right] + \\ &+ 2K|\Delta_1|^2|\Delta_2|^2 \left[ \cos(2(\phi_1 - \phi_2)) -1 \right],
\end{split}
\label{f1_charged_neutral}
\ee
with  $\rho^2= |\Delta_1|^2 + |\Delta_2|^2=1$.

For finite values of $K$, at low but finite temperatures, the system exhibits an algebraic-ordered SC state, that is destroyed at higher temperatures. However, a spontaneous symmetry breaking does occur in the $Z_2$ sector, which is associated with the two-fold degeneracy of the phase difference $\phi_{1,2}= \phi_1 - \phi_2 = \pm \pi/2$, resulting from the presence of the biquadratic Josephson term.



To obtain the phase diagram of the model \eqref{f1_charged_neutral}, and in particular identify the presence of a fermionic quadrupling condensate, 
it is necessary to assess, as a function of the parameter $K$, the two critical temperatures $T_{BKT}$ and $T_c^{Z_2}$.
For (i) $T_{BKT} > T_c^{Z_2}$, there arises a superconducting phase that preserves time-reversal symmetry; while for (ii) $T_c^{Z_2} > T_{BKT}$, a metallic state that breaks the time-reversal symmetry forms as a result of the condensation of fermion quadruplets~\cite{Bojesen2013time, Bojesen2014_phase,Grinenko2021_state, Maccari2022Effects}. The observation of a quadrupling-fermionic condensate was recently reported in the three-dimensional material Ba$_{1-x}$K$_x$Fe$_2$As$_2$ ~\cite{Grinenko2021_state}. 

 {The problem of whether a multicomponent system has a single transition or a four-fermion order is very complicated to assess, and most of the progress on such systems to date comes from large-scale numerical simulations \cite{Bojesen2013time,Bojesen2014_phase,Kuklov2008_deconfined,kuklov2006deconfined}.
Indeed, these nonsuperconducting phases are large and directly amenable for analytical arguments only in a few cases, such as two-dimensional superconductors with finite magnetic field penetration length \cite{Babaev2004_phasediagram} or systems where such order can be induced and tuned by an external magnetic field \cite{babaev2004superconductor}.}
 
 {The reason why the problem is that challenging is that beyond the mean-field approximation, the physics of the system, and therefore its phase diagram, is governed by the proliferation of topological phase excitations that mutually interact with each other.} These can be elementary vortex excitations, resulting from a phase winding in each condensate individually; composite vortices, resulting from the phase winding of both condensates around the same core; and domain walls separating regions with opposite phase differences.  The  elementary vortices $(\Delta \phi_1= \pm 2\pi, \Delta \phi_2= 0) \equiv (\pm 1, 0)$ or $(\Delta \phi_1= 0, \Delta \phi_2= \pm 2\pi) \equiv ( 0, \pm 1)$ have a phase winding in the intercomponent phase difference and hence emit a domain wall. Consequently, their proliferation restores the $Z_2$ symmetry and simultaneously destroys the superconducting state leading to the BKT superfluid-stiffness jump to zero at the critical point.
On the other hand, the proliferation of composite vortices of the kind $\pm (1,1)$ can only affect the superconducting sector, leaving the $Z_2$ symmetry broken.
Likewise, the proliferation of domain-wall excitations alone can only restore the $Z_2$ symmetry leaving the superfluid stiffness associated with the SC phase finite.  
 {The key problem is that the defects in the $U(1)$ and $Z_2$ sectors are not decoupled.
First, in contrast to the ordinary vortices  \cite{nelson1988vortex}, the composite vortices in this model consist of two spatially separated fractional vortices because of the condition  $\rho^2= |\Delta_1|^2 + |\Delta_2|^2=1$. 
Such defects carry a skyrmionic topological charge and exist also when
one softens the $\rho^2=1$ constrain \cite{garaud2015properties}. 
That implies that in contrast to conventional  vortices thermal excitations in the $U(1)$ sector also generate local
defects in the phase-difference sector. Likewise, a thermally induced Ising domain wall interacts with vortices by splitting them into two half-quanta vortices \cite{garaud2015properties}, hence disorder in the $Z_2$ sector 
may induce disorder in the $U(1)$ sector as well. As a consequence, it would in principle be incorrect to assess the critical temperatures of these two sectors by treating them separately, there are several correlation lengths and their interplay is highly nontrivial.}

\section{Monte Carlo simulations}

In this work, we address this phase transition via large-scale Monte Carlo simulations of the two-dimensional model \eqref{f1_charged_neutral}. The discrete Hamiltonian reads:
\be
\begin{split}
H = &- \sum_{i,\mu}\sum_{\alpha=1,2}| \Delta_{\alpha,i}| | \Delta_{\alpha,i +\mu} | \cos{(\phi_{\alpha,i+\mu} -\phi_{\alpha,i})} +\\&  + \sum_{i} K |\Delta_{1,i}|^2|\Delta_{2,i}|^2 \left[ \cos(2(\phi_{1,i} - \phi_{2,i})) -1 \right], 
\label{H_discrete}
\end{split}
\ee
where $\mu=\hat{x}, \hat{y}$ and $| \Delta_{1,i}|^2 + | \Delta_{2,i}|^2 =1 \,\,\, \forall i \in [0, L\times L)$.
Further details of the numerical simulations are discussed in the Supplementary Information~\cite{SM}.

\begin{figure}[b!]
    \centering
    \includegraphics[width=0.9\linewidth]{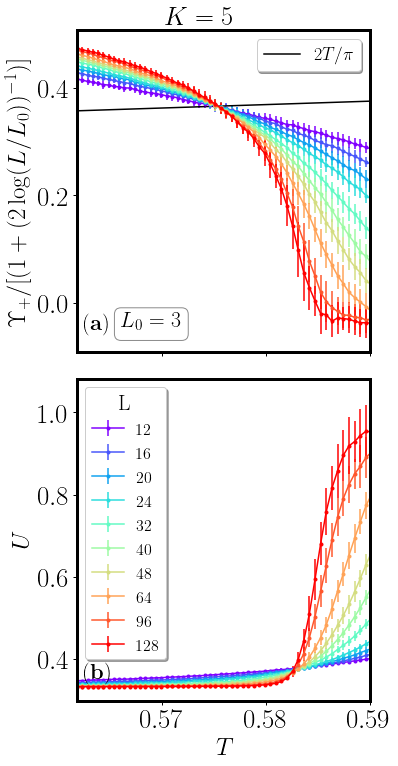}
    \caption{(a) Helicity-modulus sum $\Upsilon_+$ rescaled according to Eq.\eqref{scaling_BKT} with $L_0=3$ and (b) Binder cumulant $U$ as a function of the temperature $T$ for the case $K=5$.  We plot different values of the linear system size $L$ so as to show the two crossing points.}
    \label{crossings_K1}
\end{figure}

The BKT superconducting transition is associated with the
emergence of a finite stiffness of the phase-sum. Within the Ginzburg-Landau model Eq.\eqref{f1_rescaled_typeII}, this can be assessed by computing the helicity-modulus sum $\Upsilon^{\mu}_{+}$, defined as the linear response of the system to an infinitesimal  twist of the two phase condensates along the direction $\mu$:


\begin{equation}
\begin{split}
\Upsilon^{\mu}_{+} =\frac{1}{L^2} \frac{\partial^2 F(\{\phi'_i\}) }{\partial \delta_{\mu}^2}\Bigr|_{\delta_{\mu}=0}= \Upsilon^{\mu}_{1} + 2 \Upsilon^{\mu}_{12},
\end{split}
\label{Helicity_multicomponent1}
\end{equation}
where:

\begin{equation}
\label{Helicity1}
\begin{split}
 \Upsilon^{\mu}_{i=1,2}= \frac{1}{L^2} \Big[ \Big\langle \frac{\partial^2 H}{ \partial \delta_{\mu,i}^2}
  \Big\rangle   -\frac{1}{T} \Big\langle \left(  \frac{\partial H}{ \partial \delta_{\mu,i}} - \langle \frac{\partial H}{ \partial \delta_{\mu,i}} \rangle \right)^2  \Big\rangle  \Big]_{\delta_{\mu}=0}; 
  \end{split}
 \end{equation}
 \begin{equation}
  \label{Helicity2}
 \begin{split}
 \Upsilon^{\mu}_{12}=  -\frac{ 1 }{TL^2} \Big[  \Big\langle \frac{\partial^2 H}{ \partial \delta_{\mu,1} \partial \delta_{\mu,2} } \Big\rangle  -\langle \frac{\partial H}{ \partial \delta_{\mu,1}}\rangle \Big\langle \frac{\partial H}{ \partial \delta_{\mu,2}} \rangle   \Big\rangle \Big]_{\delta_{\mu}=0}.
 \end{split}
\end{equation}

Here, $\delta_{\mu, i}$ denotes the phase-twist parameter with respect to the i-th phase component.
Here, $L$ is the linear size of the two-dimensional system. The expectation value $\langle \dots \rangle$ is the thermal average, evaluated stochastically by the Monte-Carlo Metropolis algorithm.
In our simulations, we compute the helicity-modulus sum along $\mu=\hat{x}$. In what follows, we will simply write: $\Upsilon_{+} \equiv \Upsilon^{\hat{x}}_{+}$.

Ordinary $U(1)$ systems in two dimensions exhibit a topological phase transition driven by the unbinding of vortex-antivortex pairs ~\cite{Berezinskii1972, Kosterlitz1972, Kosterlitz1973}, which becomes entropically favorable at a finite temperature $T_{BKT}$. 
The proliferation of free vortices leads to a discontinuous vanishing of the phase stiffness, that drops to zero at $T_{BKT}$ according to the Kosterlitz-Nelson universal relation~\cite{nelson1988vortex}.

When a system undergoes a BKT phase transition ~\cite{Berezinskii1972, Kosterlitz1972, Kosterlitz1973}, the critical point can be located by finite-size scaling of the quantity~\cite{WeberMonteCarlo1988}:
\be
\Upsilon_+(\infty, T_{BKT})=\frac{\Upsilon_+(L, T_{BKT})}{1 + (2\log(L/L_0))^{-1}},
\label{scaling_BKT}
\ee
where $L_0$ is a free parameter giving the best crossing point at finite temperature (see also Supplementary information~\cite{SM} and Supplementary Fig. 1). 
 For $K=5$, the best crossing point is obtained for $L_0=3$, as shown in  Fig. \ref{crossings_K1}(a).   
Varying $K$, the value of $L_0$ varies as well. In particular, we find that $L_0$ increases with decreasing $K$ (see Supplementary Fig. 2 and Supplementary Fig. 3), leading to very pronounced finite-size effects at small $K$. This finding stems from the multi-component nature of the system.   
Indeed, in contrast to the single-component case, the BKT transition is in this case driven by the proliferation of free composite vortices, resulting from the unbinding of a pair formed by a $(1,1)$ and a $(-1,-1)$ vortex. For large values of $K$, the superconducting phases of the two condensates are essentially locked, and the model \eqref{f1_rescaled_typeII} can effectively be described by a single component. In this limit, the two elementary vortices $\pm (1,0)$ and $\pm (0,1)$ that constitute $\pm (1,1)$ composite vortices are tightly bound. 
However, for smaller values of $K$, this is no longer the case. Indeed, alongside the density-density interaction that promotes the separation of the composite vortices into their elementary constituents, in the limit $K\to 0$ the model approaches the $SU(2)$ symmetry where the composite vortices are unstable in a conventional sense.
The finite size of these composite vortices results in an increase of the finite-size effects of the whole system leading to a larger value of $L_0$.  That also suggests that standard conventional-vortex-based estimate for the BKT transition are not accurate in this limit.

To asses the $Z_2$ phase transition, we define an effective Ising order parameter $m$, related to the two possible values of $\phi_{1,2} \in [-\pi; \pi)$ via:

\be
\begin{cases}
m=+1 \,\,\,\,\, \phi_{1,2}\geq 0 \\
m=-1 \,\,\,\,\, \phi_{1,2}< 0.
\end{cases}
\ee
Finally, we extract the $Z_2$ critical temperature by means of a finite-size crossing analysis of the Binder cumulant $U$ of $m$:
\be
U=\frac{\langle m^4\rangle}{3 \langle m^2\rangle ^2}.
\label{Binder}
\ee
In the thermodynamic limit, $U$ tends to $1$ in the high-temperature phase and to $1/3$ in the low-temperature limit. The resulting crossing point for the case $K=5$ is shown in Fig. \ref{crossings_K1} (b). 
\begin{figure}[t!]
    \centering
    \includegraphics[width=\linewidth]{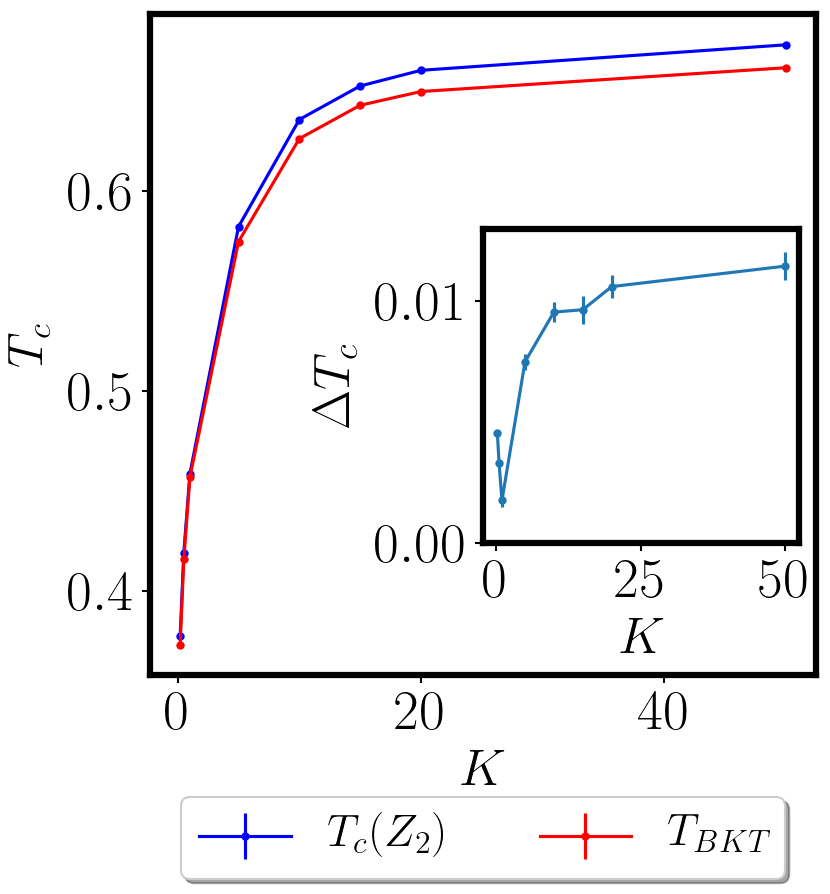}
    \caption{Phase diagram of the model Eq. \eqref{f1_rescaled_typeII} as a function of the coupling $K$. For any finite value of $K$, the 
    BKT and $Z_2$ transitions are found to be separated, with $T_{BKT} > T^{Z_2}_c$.  In the inset, the size of the observed splitting $\Delta T_c= T_{BKT} - T^{Z_2}_c$ is reported as a function of $K$. The largest splitting is found for smaller values of $K$, while in the limit of $K\to \infty$ it saturates to a finite value. }
    \label{phase_diagram}
\end{figure}

The phase diagram obtained via this numerical study is shown in Fig. \ref{phase_diagram}. Details on the finite-size scaling of the two critical temperatures can be found in the Supplementary Information~\cite{SM} (Supplementary Fig. 4-6). Our results reveal that for any finite value of $K$ we considered,  the system has a fermionic quadrupling state that breaks time-reversal symmetry. The range of temperatures where this phase appears (see the inset of Fig. \ref{phase_diagram} ) is larger for large values of $K$ and saturates to a finite value in the limit $K\to \infty$.
%
The presence of a   lattice provides a minimum size for the domain wall between two different chiralities. Consequently, the energy cost of such topological defects saturates to a finite value in the limit $K \to \infty$, resulting in a saturation of the critical temperature associated with the $Z_2$ transition.
In  contrast to the previously studied conventional multiband models, in the limit of a very small intercomponent  coupling $K$, the two transitions do not merge and we observe a relative increase of $\Delta T_c$ for $K<1$. We argue that this increase is related to the symmetry of the model in the limit $K \to 0$ that, for the model derived in \cite{chichinadze2020NematicScTBG}, is $SU(2)$, rather than $U(1)\times U(1)$ symmetry as for the case of $s+is$ superconductors.  In two dimensions, $SU(2)$-symmetric systems exhibit no long-range or quasi-long-range order.  
\\
Obtaining a significant fermion quadrupling phase in the case when $U(1)\times U(1)$ symmetry is explicitly broken to $U(1)\times Z_2$ generally requires a very strong symmetry-breaking Josephson term \cite{Bojesen2013time}. By contrast, in the TBG model considered in this work,  { a noticeable quadrupling phase remains even if the term that breaks the $Z_2$ symmetry is small. In all the considered cases however the positions of the critical points are very correlated, and the difference in critical temperatures is of the order of $1 \%$ signalling that fluctuations in the $U(1)$ and $Z_2$ sectors are nontrivially coupled.} Therefore, the phase diagram of this or similar models cannot be accurately determined by treating the $U (1)$ and $Z_2$ sectors separately.

\section{Conclusions}

In conclusion,  
we studied a Ginzburg-Landau model 
derived in connection with  twisted bilayer graphene  \cite{chichinadze2020NematicScTBG} at low temperatures. 
We have shown that this class of models
can host
a fermion quadrupling phase above the critical temperature of the superconducting phase. 
This phase is more robust than its counterpart in the previously studied class of models \cite{Bojesen2013time,Bojesen2014_phase,
Haugen2021first,zeng2021phase} and extends to all finite values of the coupling parameter $K$ considered but is still relatively small. Nonetheless
%
these findings indicate that magic-angle twisted bilayer graphene can be an especially promising platform for realizing and observing fermion quadrupling order.

While the temperature range  associated with fermion quadrupling is small, it is likely to be larger in real systems. The coupling to a vector potential --omitted in this work-- reduces the energy of composite vortices, thus reducing the temperature of the onset of superconductivity  {so that a finite diamagnetism could lead to a significant fermion quadrupling phase.}
More importantly, the robustness of this phase in the model we studied
suggests that its size can be amplified
by applying a transverse magnetic field
%
%
\cite{babaev2004superconductor,Grinenko2021_state}.

The fermion quadrupling state can be identified via a combination of thermal and electrical transport measurements, analogous to those performed in \cite{Grinenko2021_state}. 
The effective model of the $Z_2$ quadrupling state \cite{garaud2022effective} suggests that signatures of a $Z_2$ broken symmetry above the critical temperature can be detected via magnetic probes. Skyrmion excitations \cite{garaud2022effective} or spontaneous magnetic fields can indeed appear in the presence of local strain, obtained by imposing local pressure or by local heating \cite{garaud2016thermoelectric,silaev2015unconventional} in combination with local magnetic probes. 
Another route to probe this state in twisted bilayer graphene is through collective modes \cite{Carlstrom2011_lengthscales,lin2012massless, Stanev2012,poniatowski2022spectroscopic}.
The inter-component collective modes, indeed, only depend on the relative phases and relative densities of the two components and thus they should survive in the non-superconducting state with broken time-reversal symmetry.


 \section*{Acknowledgements}
 We thank Mats Wallin, Jack Lidmar and Pablo Jarillo-Herrero for useful discussions.
The simulations were performed on resources provided by the Swedish National Infrastructure for Computing (SNIC) at the National
Supercomputer Center at Link\"oping, Sweden. I.M. acknowledges the Carl Trygger foundation through grant
number CTS 20:75. E.B. is supported by the Swedish
Research Council Grants 2016-06122, 2018-03659. J.C. is supported by the Swedish
Research Council Grant 2018-03882.

\bibliography{weston}

\end{document}